\documentclass[%
 preprint,
 amsmath,amssymb,
 aps,
]{revtex4-1}

\usepackage{graphicx}
\usepackage{dcolumn}
\usepackage{bm}
\usepackage[mathlines]{lineno}
\usepackage[utf8]{inputenc} 
\usepackage[normalem]{ulem}

\usepackage{amsmath, graphicx, enumitem, multirow, mathtools}
\newcommand{\zmed}{$z_{med}$}
\newcommand{\zmin}{$z_{min}$}

\begin{document}


\title{Technological novelty profile and invention's future impact}

\author{Daniel Kim}
 \affiliation{Natural Science Research Institute, Korea Advanced Institute of Science and Technology, 291 Daehak-ro, Daejeon 34141, Republic of Korea}
 \affiliation{Santa Fe Institute, 1399 Hyde Park Road, Santa Fe, NM 87501, USA}

\author{Daniel Burkhardt Cerigo}
 \affiliation{Institute for New Economic Thinking at the Oxford Martin School, Oxford OX2 6ED, United Kingdom}
 \affiliation{Santa Fe Institute, 1399 Hyde Park Road, Santa Fe, NM 87501, USA}

\author{Hawoong Jeong}%
 \affiliation{Department of Physics, Korea Advanced Institute of Science and Technology, 291 Daehak-ro, Daejeon 34141, Republic of Korea}
 \affiliation{Institute for the BioCentury, Korea Advanced Institute of Science and Technology, 291 Daehak-ro, Daejeon 34141, Republic of Korea}
 \affiliation{Asia Pacific Center for Theoretical Physics, 67 Cheongam-ro, Pohang 37673, Republic of Korea}

\author{Hyejin Youn}%
 \email{youn@math.ox.ac.uk}
 \affiliation{Institute for New Economic Thinking at the Oxford Martin School, Oxford OX2 6ED, United Kingdom}
 \affiliation{Santa Fe Institute, 1399 Hyde Park Road, Santa Fe, NM 87501, USA}
 \affiliation{Mathematical Institute, University of Oxford, Oxford OX2 6GG, United Kingdom}

\date{\today}

\begin{abstract}
We consider inventions as novel combinations of existing technological
capabilities. Patent data allow us to explicitly identify such combinatorial
processes in invention activities \cite{HYoun2015}. 
Unconsidered in the previous research, not
every new combination is novel to the same extent. 
Some combinations are naturally anticipated based on patent
activities in the past or mere random choices, 
and some appear to deviate exceptionally
from existing invention pathways. We calculate a relative likelihood 
that each pair of classification codes is put together at random, 
and a deviation from the empirical observation so as to assess 
the overall novelty (or conventionality) that the patent brings forth at each year. 
An invention is considered as unconventional if a pair of codes therein is 
unlikely to be used together given the statistics in the past. 
Temporal evolution of the distribution indicates that the patenting activities become
more conventional with occasional cross-over combinations. 
Our analyses show that patents introducing novelty on top of the conventional units 
would receive higher citations, and hence have higher impact. 
\end{abstract}


\maketitle


\section*{1 Introduction}
A new idea that advances science and technology is commonly 
recognised as an important source of wealth creation, economic growth, 
and societal change~\cite{Schumpeter1939,Nelson1993}. 
The steam engine, transistor and lithium ion battery are all such examples. 
Therefore, it is of no surprise that understanding the dynamics of generation of new ideas sits at the centre of many disciplines~\cite{Page2008,Evans2011, Wagner2014, Corominas-Murtra2013,Uzzi2013,Malva2014, Strumsky2015, JYun2015, Farmer2015,JWang2015,Sinatra2015,ONeale2012,Bettencourt2014,HYoun2014}. 

With an increasing volume of electronic corpora available online, research on the systems of science and technology, once considered to be a domain of humanities, social sciences and economics, has expanded its realm to be a subject of data science~\cite{Evans2011}. The growing empirical literature in this respect is to identify the process of publication~\cite{Calcagno2012}, to utilise Google n-gram to characterise scientific evolution~\cite{JYun2015}, to delineate the boundary of science~\cite{Rosvall2008,Borner2015}, and to predict the future impact of scientific papers~\cite{YHEom2011,DWang2013} and authors~\cite{Deville2014,Petersen2014}. 

In the case of inventive activities, Youn and co-workers availed themselves of technology codes, 
classified by United States Patent and Trademark Office (USPTO), as countable units 
to identify the underlying dynamics of inventions as combinatorial processes 
in a comprehensive and explicit way~\cite{HYoun2015}. 
In this way, an invention yields either a new unit of technological capability, 
a new way of combining the already existing units making an innovative function,
or a refinement of existing combinations.
The rate at which the new combinations are introduced has been \textit{invariant} over two centuries,
implying that a combinatorial process is the nature of invention. 

Building on these previous findings, we delve into the temporal evolution of this combinatorial 
process using U.S. patent data from 1836 to 2014.
We first describe the data structure, and elaborate our method to quantify technological novelty scores.
We then show how the z-scores of the pairings within an invention would affect its future impact. 
Finally we will discuss the implications of our findings. 


\section*{2 Data}
The U.S. patent records began at July 31, 1790 with Samuel Hopkins' patent on pot ash~\cite{Hopkins1790}. 
Since then, there has been almost ten million inventions granted over two hundred years~\cite{USPTObulk}. 
We only consider utility patents, which are those that pertain to new and useful inventions, omitting design and plant patents for instance~\cite{USPdesc}. There are 8,884,909 utility patents up to December 2014, taking up almost 90\% of the documents. 
Among them, we analyse patents that have two or more codes (80.3\%). 

In order for examiners to efficiently search for relevant prior arts, the U.S. patent office encode 
salient technological capabilities into six-position alphanumeric codes. Every patent is then tagged with a combination of codes that represent the technologies involved in the invention~\cite{USPCSoverview}. 
The classification codes are created in a nested structure: 473 classes at the highest level and 168,743 codes at the lowest (most detailed) level. 
These low-level codes (`codes' from herein) can lie on different levels of the hierarchy tree; some classes have deeper branching than others. 

We used patent citation data provided by National Bureau of Economic Research (NBER) and considered patents' citations as a measure of their impact~\cite{Hall2001, NBERdata}. The citation data span only 31 years, from 1976 to 2006 unlike the co-occurrence data covering almost two hundred years. In order to cover as large data as possible, we use the co-occurrence data, spanning over one hundred years, for the section 4.1, and use NBER data when citation information is needed in the section 4.2. 
In order to control 
the temporal effect on citation volume, we use the citation number only up to the first five years after publication. It is also known that only recent citation works well in prediction of future impact~\cite{Benson2015}. This leaves us data spanning 36 years (from 1976 to 2001)~\cite{Uzzi2013, Valverde2007}.

\section*{3 Methods}

We aim to assess the novelty of technological constituents in each patent, 
and then compare aspects of this novelty to the patent's impact. 
We measure how technology codes are combined in the empirical data and 
compare the observed combination to what would be expected if the combinations were randomly configured.
In this way, we can discern recurring themes within invention space 
and also those combinations that are unconventional or novel. 
These features, measured by well established standard scores, are related 
to patent future impact.

\section*{3.1 Standard scores (z-scores) of code pairs}

The patent data $P$ can be represented by a collection of sets of classification codes, where each set corresponds to an individual patent and contains its classification codes. The z-score for a pair of codes, $\alpha$ and $\beta$ is expressed as:
\begin{equation}
z_{\alpha \beta} = \frac{o_{\alpha \beta} - \mu_{\alpha \beta}}{\sigma_{\alpha \beta}}
\end{equation}
where $o_{\alpha \beta}$ is the observed number of times the code $\alpha$ appears together with 
$\beta$ within a patent (a set) within the actual data. $\mu_{\alpha \beta}$ and $\sigma_{\alpha \beta}$ are the expected co-occurrences of the codes
and its standard deviation, derived from a null model of the data which randomises code arrangement while preserving code usage and number of patents within the data (the section 3.2 provides the detail).

The observed co-occurrences $o_{\alpha \beta}$ in the patent record is compared with $\mu_{\alpha \beta}$. 
If the two codes appear together more often than expected, then Eq. (1) results in a positive value, or if they are rarely paired within a patent relative to their expected occurrences then their z-score is negative.
The degree to which the deviation is significant is derived by normalising the value by the expected standard deviation $\sigma$~\cite{Uzzi2013,Malva2014,Tibely2013,YYAhn2011}. 
We can thus associate high z-scores with very typical code pairing, and conversely, a negative z-score is indicative of an atypical or novel pairing of codes.

\section*{3.2 Expected co-occurrences} 
The null model acts as the baseline by which we deem an aspect of the data to have statistical significance, beyond what would occur by random, or with no underlying pattern or law. 
The aspect in consideration is the arrangement of the codes between the patents.
The premise of the null model is that each of these arrangements of codes is equally likely. From these possible arrangements the expected pairing counts can be computed.

Consider codes $\alpha$ and $\beta$, with the number of occurrences $n_\alpha$ and $n_\beta$ within the set of patents $P$. Noting that patents cannot be classified with the same code twice, the number of possible configurations of the $\alpha$ and $\beta$ into the $|P|$ possible patents is ${|P|\choose n_\alpha}{|P|\choose n_\beta}$. Now consider those arrangements which contain exactly $x$ co-occurrences, within a patent, of $\alpha$ and $\beta$. There are ${|P|\choose n_\alpha}{n_\alpha\choose x}{|P|-n_\alpha\choose n_\beta-x}$ possible configurations; first distribution the $\alpha$ into $|P|$, then $x$ of the $\beta$ into those $n_\alpha$ patents already assigned an $\alpha$, finally distribute the remaining $\beta$ into the patents without an $\alpha$. Thus giving a hypergeometric probability distribution for the number of co-occurrences:
\begin{equation}
p(o_{\alpha \beta}=x) = \frac{ { {n_\alpha}\choose x} { {|P|-n_\alpha}\choose{n_\beta-x}}} {{  {|P|}\choose{n_\beta}}}
\end{equation}


thus the expected number of patents that have both $\alpha$ and $\beta$ is: 
\begin{equation}
\mu_{\alpha \beta} = \frac{{n_\alpha}{n_\beta}}{|P|}
\end{equation}
and
the variance of $\mu_{\alpha \beta}$ is:
\begin{equation}
\sigma^{2}_{\alpha \beta} = \mu_{\alpha \beta} \left(1-\frac{n_\alpha}{|P|}\right) \left(\frac{|P|-n_\beta}{|P|-1}\right)
\end{equation}


\section*{3.3 Incorporating temporal evolution}

As new technologies become successful, so they may subsequently 
become established areas of inventive activity. 
Following z-scores in time 
allows us to observe the case where an invention may have been exceptionally novel in its time of creation, but its novelty 
would `wash-out' with many similar inventions subsequently follow it over time.

To capture this time variance we consider z-scores specific to time-ordered subsets of the entire data. 
We choose cumulatively increasing subsets in yearly steps,
letting $P(t)$ be the sub-collection of patents up to the year $t$ in $P$. So $P(2000)$ contains all patents issue up to the year 2000, and the z-scores calculated using this set are specific to this year. Thus for a given year, the newly added patents' z-scores 
are discerned based on all the patents that precede them, and the older patents' z-scores 
continue to evolve and change based on subsequently issued inventions.

\section*{3.4 A schematic for three cases: atypical, typical and neutral}

We provide a schematic to aid in our understanding of how atypicality 
embedded in code combinations is captured and expressed by 
z-score measure. 

Suppose 40 inventions $P$ at time $t$, indexed by its entering order $i$. 
Each invention is expressed as a combination of codes:
\begin{align*}
    P (t) = & \{P_1, P_2, P_3, \cdots, P_{40}\} \\
          = & \{ \{A,C\}\times 19, \{C,D\}, \{B,E,F\} \times 4, \\
            & \{B, E\} \times 6, \{E, F\} \times 5, \{B,F\} \times 5 \}
\end{align*}

\begin{figure}[t]
  \begin{center} 
  \includegraphics[width=0.9\textwidth]{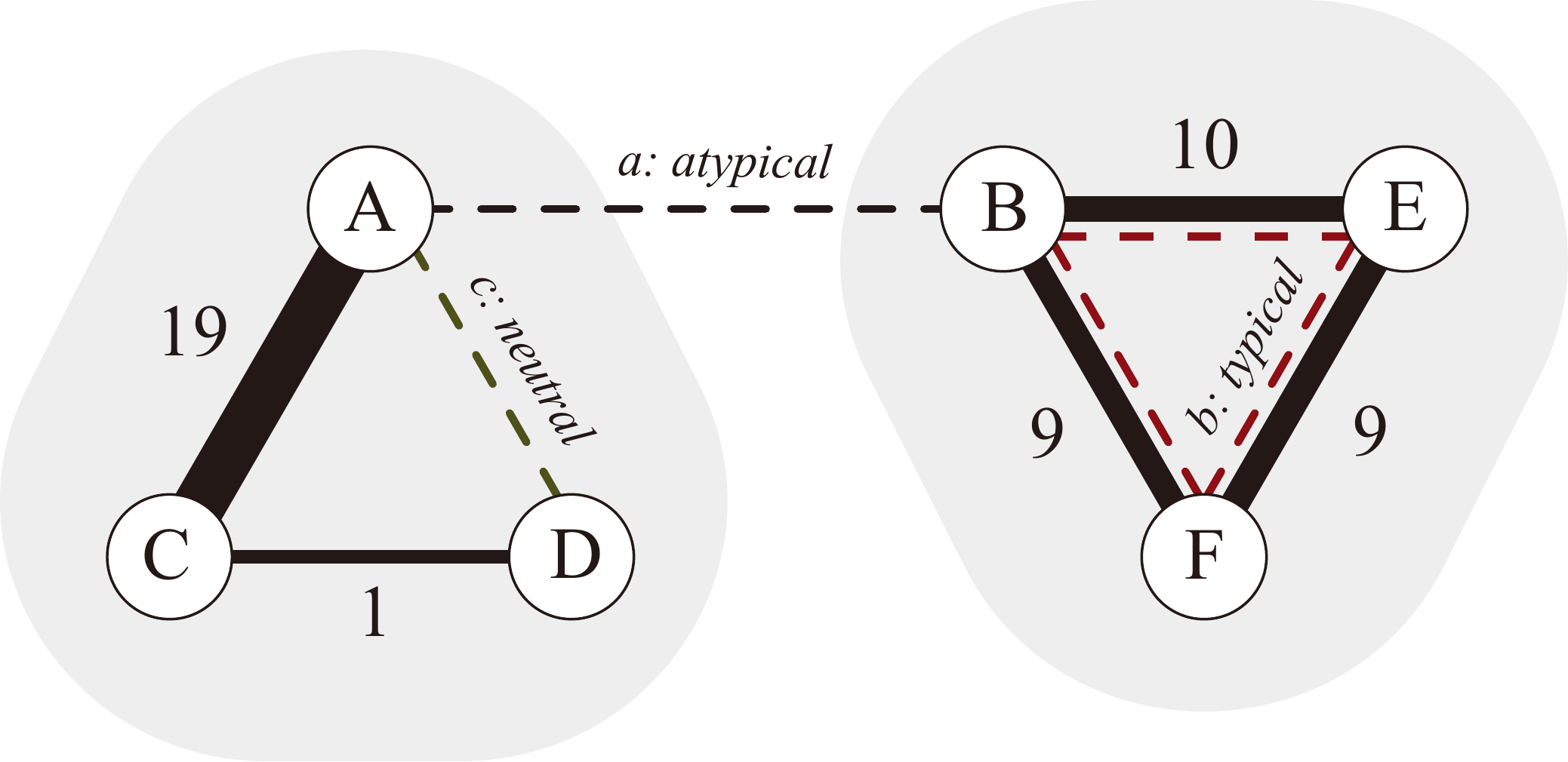}
  \caption{ The shades guide two partitioned segments in a network representation. 
   The solid lines denote the existing weighted pairs up to $t$ while 
   the dotted lines ($a$, $b$, $c$) possible additions introduced by
   newly created combinations (a set of codes used in a patent) at $t+1$.     
   Depending on where a new combination is overlaid in the network structure, 
   the invention at $t+1$ consists atypical, typical or neutral pairs relative to the random chance.}
\label{fig:schematic} 
  \end{center}
\end{figure}

Figure~\ref{fig:schematic} illustrates the collection of patents at $t$, $P(t)$, represented as a network structure 
where pairwise combinations are represented as weighted links (solid lines). 
We then consider three cases where a new patent $P(t+1)$ arrives with two codes, 
that is, (i) $P(t+1)-P(t) = \Delta P(t) \supset \{A,B\}$, denoted as a black dash line 
(ii) $\Delta P(t) \supset \{B,E\}$, as a red dash line, or (iii) $\Delta P(t) \supset \{A,D\}$, 
as a green dashed line. Simply put, links are solid when they are present at time $t$, and dashed 
when they are added at time $t+1$. 

In the case (i), the link $a$ bridges the two most frequently used codes $A$ and $B$ that are
yet combined together until time $t$. We therefore find 
the appearance of link $a$ {\it atypical} given the current statistics, 
and naturally expect a {\it negative} z-score. Indeed, calculated $z_a$ exhibits a negative value, 
that is, $-4.3$, with $\mu_{AB} = 7.8$, and  $\sigma_{AB}=1.6$ in the Eq. (1). 
Note that the frequency of $A$ and $B$ are, respectively, $n_A(t+1) = 20 $ and  $n_B(t+1) = 16$. 
On the other hand, the link $b$ reinforces the existing pair that are already well connected, or
established, hence, becoming a {\it convention}, yielding a positive z-score, $3$. This indicates 
that they are combined more than expected by three times of standard deviation derived by the random choices.
Finally, the link $c$ yields a statistically {\it neutral} pair, around zero, indicating the occurrence is indistinguishable
from the random configurations. 

\section*{3.5 Coarse-graining over classification codes}

The classification codes are created in a nested structure (hierarchical tree). The number of classes at the highest level is 473, 16,087 subclasses at the second level down, and 168,743 codes at all the way down to the lowest level.  
The codes can be extremely detailed in their content, making results pertaining to specific codes very narrow in scope, or it can be quite broad. 
For example, the class 257, ``ACTIVE SOLID-STATE DEVICES'', has the longest depth up to 16 at the end of which ``Floating gate layer used for peripheral FET (EPO)'' and ``Floating gate dielectric layer used for peripheral FET'', while the class 245, ``Wire fabrics and structure'', has the shortest depth up to 2 at the end of which ``Chain'' and ``Coil''.
As shown in the above examples, the level of differentiation for two codes in a class can be qualitatively different according to the depth and classes.
To create a consistent level of detail in the analysis, and gain broader and more intelligible insights, we coarse-grain over the codes to look at pairings at the highest and the second highest level~\cite{McNamee2013}.  




We consider each patent as a combination of classes -- the highest level. We also consider them as a combination of subclasses, i.e at one level below the class level, to gain insight in a slightly more fine-grained technology space, as well as for comparison with the class level results. At the code level a patent is never assigned the same code twice, and so no self-pairing is possible. This is not the case for class and subclasses, as multiple codes from the same class are often assigned to the same patent. 

Rather than coarse-grain over the data before the analysis, we first preserve its detailed structure at the code level, only after which we coarse-grain over the resultant statistics to gain z-scores for class and subclass pairings. 
Let $\alpha_i$ and $\beta_j$ represent a code each, where $\alpha$ and $\beta$ denote the class and $i$ and $j$ represent the rest of the code, specifying the low level detail. Calculating the observed co-occurrences $o_{\alpha\beta}$ and expected co-occurrences $\mu_{\alpha\beta}$ of a class pair is carried out by summing over all code pairings that result in the considered class pairing. Hence for the observed co-occurrences:
\begin{equation}
    o_{\alpha\beta}=(1-\frac{\delta_{\alpha\beta}}{2})\sum_{i,j}o_{\alpha_i\beta_j}
\end{equation}
where the bracketed term accounts for double counting when the classes considered are the same. Similarly for the expected co-occurrences:
\begin{equation}
    \mu_{\alpha\beta}=(1-\frac{\delta_{\alpha\beta}}{2})\sum_{i,j}\mu_{\alpha_i\beta_j}
\end{equation}
and the variance:
\begin{equation}
    \sigma^2_{\alpha\beta}=(1-\frac{\delta_{\alpha\beta}}{2})\sum_{i,j}\sigma^2_{\alpha_i\beta_j}
\end{equation}
from which using these the class pair z-score can be calculated.


In addition, unless we use the above coarse-graining, the code usages/frequencies mostly can be far smaller than the number of patents up to $t$, $|P(t)|$, because these usage values follow a power-law. That means an actual co-occurrence value $o_{\alpha\beta}$ between two different codes, $\alpha$ and $\beta$, can be much larger than expected co-occurrence, $\mu_{\alpha\beta}$, between $\alpha$ and $\beta$ though $o_{\alpha\beta}=1$. Thus, using fine-grained codes mostly gives us positive z-scores and we cannot obtain the novel, but the conventional. 

\section*{4 Results}
Technological constituents of a patent are translated into a set of pairwise z-scores that characterise its novelty or typicality. 
In this way, we are able to capture how inventors combine technological units 
by analysing the summary statistics.

In the following sections, we will look at the distribution of z-scores derived from the entire set of patents, 
and compare it with that of newly created combinations, in each year.  
Then we will relate the observed compositional features of an invention at the time of its creation 
to its future impact. All analysis is carried out at both the class and subclass level (one level down from the class) to ensure that our findings and insights are persistent across different levels of detail.


\section*{4.1 Decomposition of new combinations}\label{sec:new_pair_comb}
It has been shown that the rate at which inventors create new combinations is invariant, 
and that they do so more often than not~\cite{HYoun2015}. 
This result alludes to a ceaseless introduction of new ways of combining technological units, 
and thereby a constant reshaping of technology space. By just considering the number of new combinations occurring, 
the dynamics of novelty creation looks temporally independent, which conforms with the possibility that new inventions occur at random.

On the contrary, a new combination is not simply concocted by randomly choosing technological units, 
but, however novel it may be, it is either built on the existing body of knowledge accumulated, or discovered by 
 the expansion of the adjacent possible in the technology space~\cite{Tria2014, Arthur2011, Kauffman1996, Alstott2015}. 
The new combination may not be composed of an entirely novel membership~\cite{Strumsky2015}. 
It may contain a set of codes  that have been frequently combined, such that they can be considered as an established unit, or building block~\cite{Hidalgo2009}.  
Therefore, binary classification---a combination can be either absolutely novel if it was previously unseen, 
or otherwise not novel---misses the subtleties of an invention's novelty by lacking the complexity to capture it in any detail.
For instance, a combination with small novel addition to the conventional subset as unique. 

We decompose combinations into a novelty profile in terms of pairwise z-scores, 
and assess the extent to which the multiple aspects of a combination are novel or conventional in more detail~\cite{Uzzi2013, Malva2014}.
In this way, a new combination can both reinforce 
the current technological conventions, and introduce new ways of combining codes. 
As elaborated in the Method section, the z-scores are a means to compare the observed occurrences to the random counterpart.

\begin{figure}[h!]
    \includegraphics[width=0.9\columnwidth]{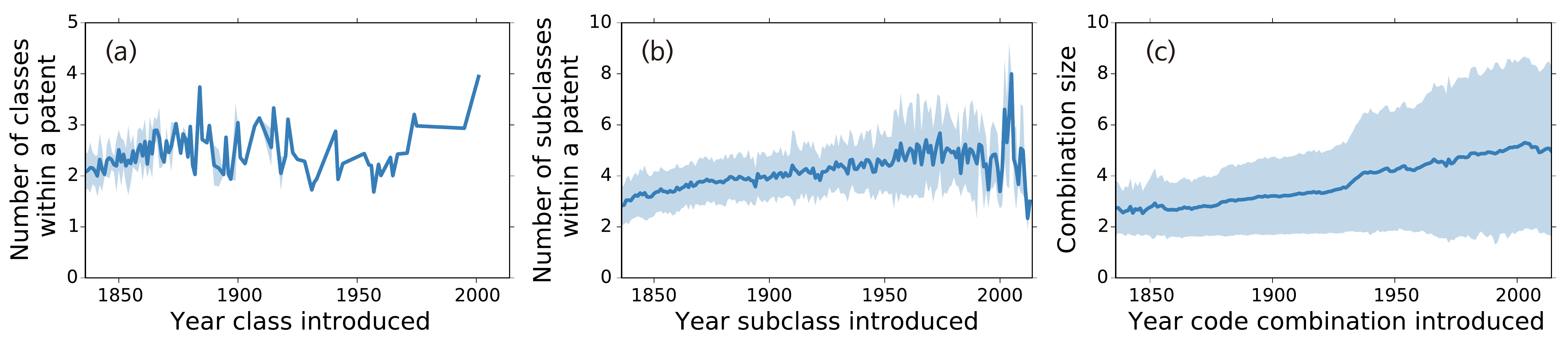}
    \caption{\textbf{Basic statistics.} 
    (a--c) each x-axis indicates year when (a) a class, (b) a subclass, or (c) a code combination was introduced. 
    (a--b) each y-axis represents the average number of (a) classes and (b) subclasses combined in those patents containing a new code. 
    (c) y-axis displays the number of codes in a new combination. Shaded area denote standard deviation.}
    \label{fig:basic_stat}
  \end{figure}

Although there are no limits to how many codes may be assigned to a patent,
Figure~\ref{fig:basic_stat} shows the number of codes in a patent hovers around three to four in average with a tendency that 
the size increases in time, 
indicating parsimonious code usages. 
Every pair within a combination is then assigned a z-score (see, the Method section), 
and the composition of an invention can be captured by three statistics: 
its median $z_{med}$ and minimum $z_{min}$, and the difference between the two 
$\Delta z \equiv z_{med}-z_{min}$~\cite{Malva2014}.  
The $z_{med}$ indicates the degree to which the main body of a patent conforms to technological conventionality,
while $z_{min}$ indicates the extent to which the invention contains an element which is novel 
when combined with the patents other parts.
The difference between the two $\Delta z$ captures aspects of both in a single measure; whether the patent has a conventional core \textit{and} a novel addition.

\begin{figure}[h!]
    \includegraphics[width=0.9\columnwidth]{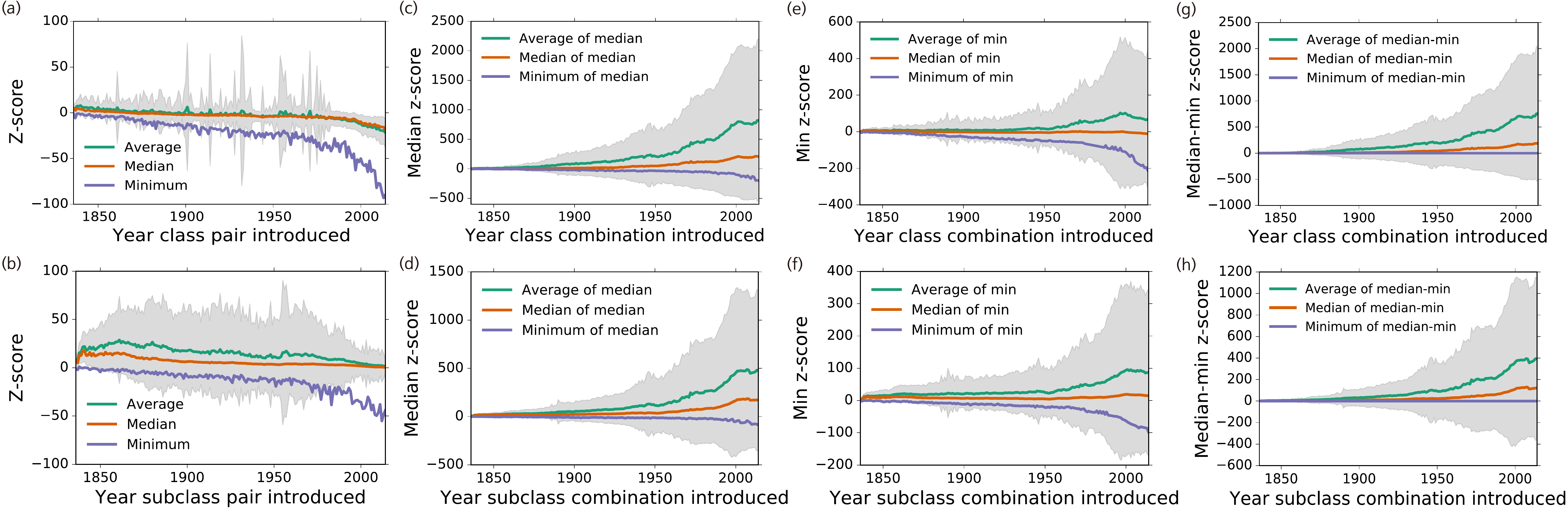}
    \caption{\textbf{Newly introduced pairs and combinations versus z-score. } 
    (a) and (b) show z-score of a new pair when it was firstly combined.
    (c--h) 
    each x-axis indicates a year that the new combination was firstly used, and 
    each y-axis represents $z_{med}$ (c--d), $z_{min}$ (e--f), and $z_{med}-z_{min}$ (g--h). 
    (a), (c), (e), and (g) are for classes. 
    (b), (d), (f), and (h) are for subclasses. 
    Shaded regions represent standard deviation.}
    \label{fig:first_year_introduced_vs_z-score_pair_combination}
  \end{figure}

Figure~\ref{fig:first_year_introduced_vs_z-score_pair_combination} (a) shows z-score of new class pairings, 
in which it is seen that their average z-score remain zero, that is, indistinguishable from the random incidence on average, 
and then gradually become negative after 1980.  
This implies, new class pairings neither strengthen nor join any modular structures of the class network 
when they firstly appear, and then new atypical class pairings after 1980 gradually join two different technological domains.
In addition, new class pairs gradually being atypical may dispute a claim that 1880s was more innovative than now~\cite{Smil2015}.

When a new pair, or combination was introduced, 
it is normally the case that its z-score is negative, or neutral as is shown in Fig. 1. 
Occasionally however, there is also a case where codes involved in the combination have rarely been used, 
that the expected co-occurrences, $\mu$, and the standard deviation, $\sigma$, is relatively low. 

We then capture the compositional features of how newly created combination by these two summary statistics of each year. 
Figure~\ref{fig:first_year_introduced_vs_z-score_pair_combination} (c) and (e) show $z_{med}$, $z_{min}$, and $\Delta z$. It shows that these z-scores steadily grow over time in a more or less margin of increase. The new combinations mostly 
contain the conventional pairings. Additionally, the gap between $z_{med}$ and $z_{min}$ increases (Fig. 3 (g) and (h)). 
In other words, a new combination becomes both more conventionality and non-conventionality.

\begin{figure}[h!]
    \includegraphics[width=0.9\columnwidth]{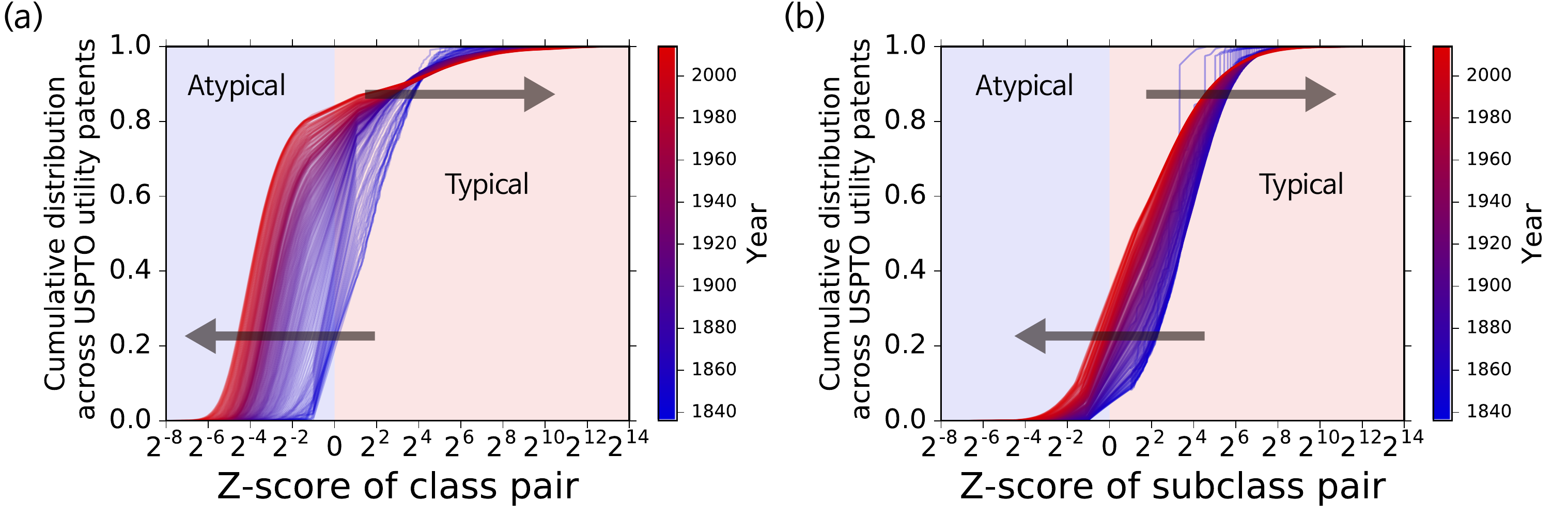}
    \caption{\textbf{Cumulative distributions of z-scores over year.} 
    (a) for class pairs. 
    (b) for subclass pairs. }
    \label{fig:cumulative_distributions}
  \end{figure}

\begin{figure}[h!]
    \includegraphics[width=0.9\columnwidth]{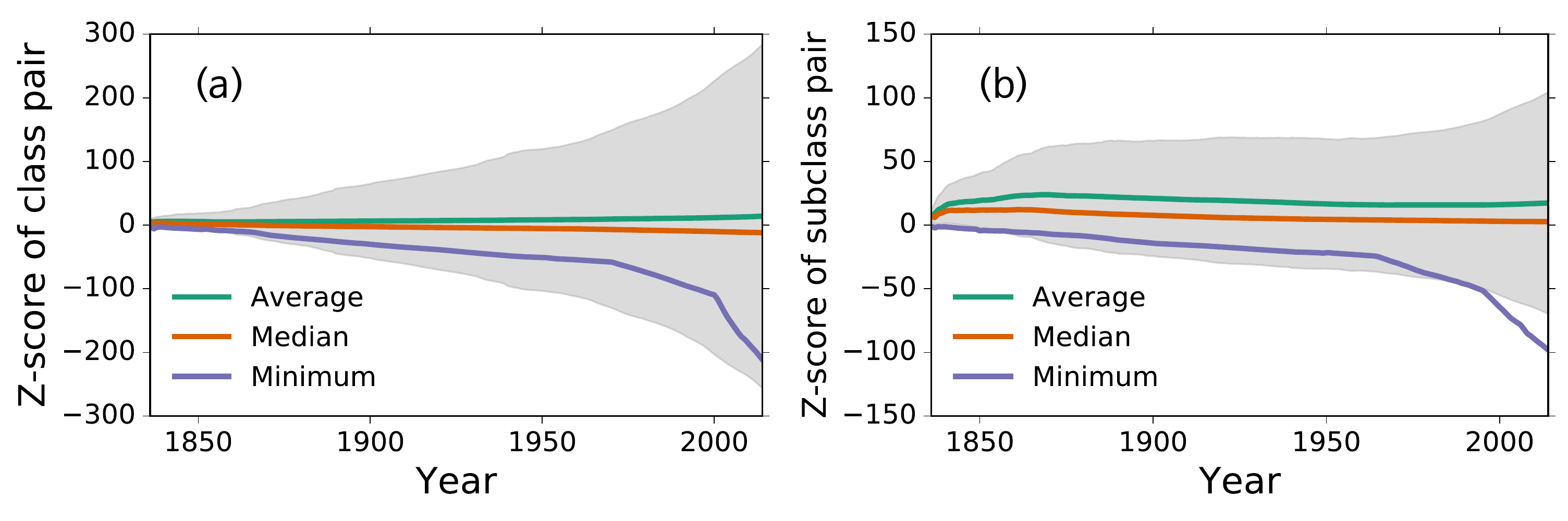}
    \caption{\textbf{Statistical properties of cumulative distributions of z-scores over year.} 
    (a) for class pairs. 
    (b) for subclass pairs. 
    Shaded regions represent standard deviation.}
    \label{fig:quantification_of_cumulative_distributions}
  \end{figure}

We now look at how these compositional features of new pair and new combinations reshape the landscape. 
We characterising this phenomena by analysing the distribution of z-scores for entire pairs 
accumulated up to the year. 
Figure~\ref{fig:cumulative_distributions} shows the cumulative distribution of z-scores for every year from 1836 to 2014, respectively denoted by a colour scale (from blue to red). 
Broadening of this distribution across time indicates that the network is becoming more ingrained, with increasingly highly connected subsets of codes, hence higher z-scores,
while pairs that span between these conventional units are thus increasingly perceived as atypical.
Thus if two codes are used together more often than random expectation, it is probable that they are used together again. 
This broadening is also explicitly shown in the Fig.~\ref{fig:quantification_of_cumulative_distributions} where the standard deviation (grey shade) widens over time and the minimum z-scores of the year cohort becomes increasingly negative,
especially around the recent decades. 

\section*{4.2 Compositional features predicting future impact}\label{sec:future_impact} 

Predicting which new inventions will have a high impact is an obviously wanted goal, both for attempting to predict profitability, but also as a signifier of future societal changes cause by new technologies~\cite{Farmer2015,Benson2015}.
Further to just assessing new inventions, an understanding of the qualities related to invention impact enables one to optimise their inventive strategy to maximise such qualities.
We show that a patent's success is predictable using its novelty profile. 

As we discussed in the previous sections, an invention is interpretable as pairwise z-scores, quantifying statistically significance of code pairings in inventing activities.
Built on the previous research, suggesting that the compositional feature is key to a patent having a high impact, we delve into the temporal dynamics of this relationship given our patent records~\cite{Uzzi2013, Malva2014}.

We define high impact inventions as those patents in the upper 5th percentile\footnote{The upper 5th percentile can be considered as a statistical significance with $p$ values of 0.05.}, within each year, of citations gained within 5 years from publication~\cite{Benson2015,Valverde2007}. 
We categorise the patents accordingly: whether (i) $z_{med}$ of a patent belongs to either the top quartile $z_{med}$ of a year, middle half, or bottom quartile, (ii) similarly $z_{min}$ in the top quartile, middle half, or bottom, and (iii) $\Delta z$ in the top, middle, or the bottom.
We abbreviated top quartile to high, middle to mid, and bottom to low.

\begin{figure}[h!]
    \includegraphics[width=0.9\columnwidth]{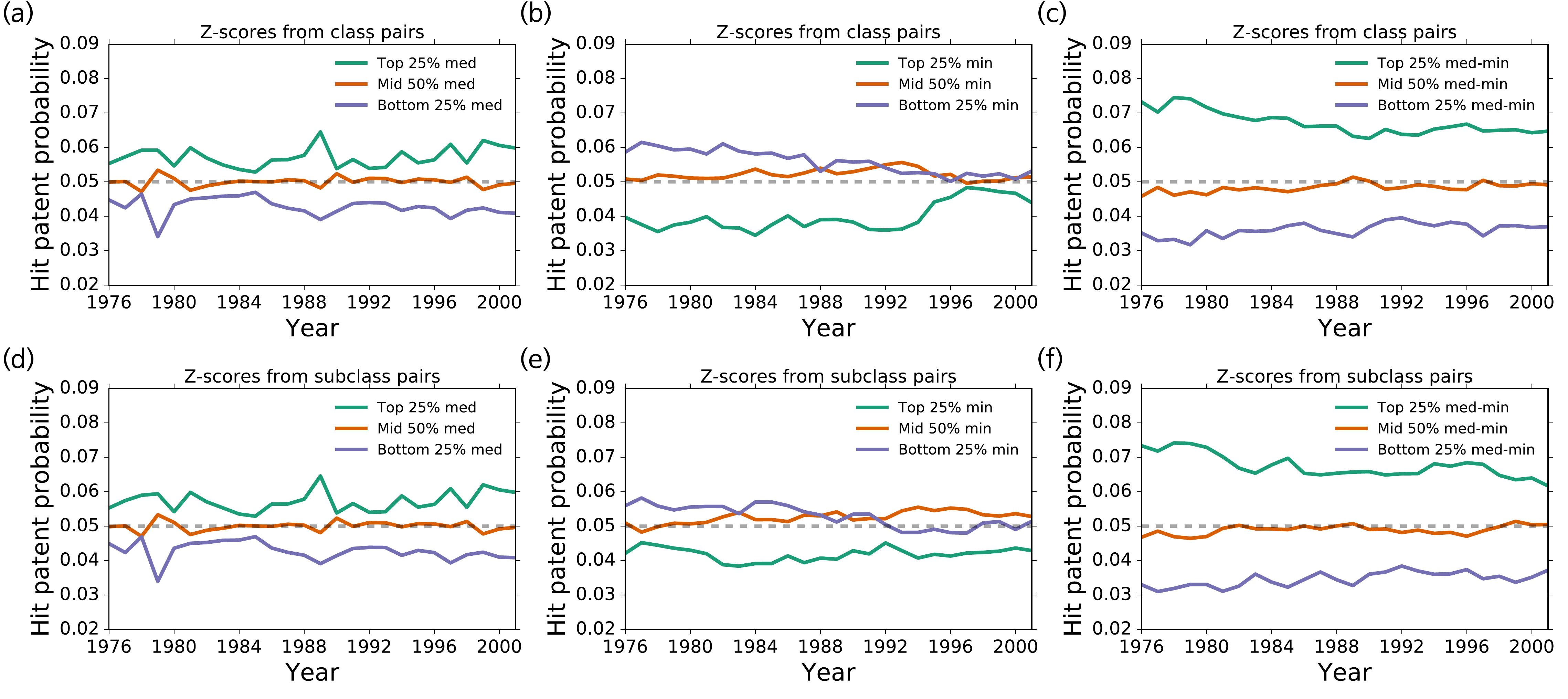}
    \caption{\textbf{Year versus ``hit'' patent probability.} 
    (a--f) show the probability that a patent is in top 5th percentile by citations among the grant year cohort. 
    Patents are partitioned into three categories by the top and the bottom quartile, and in between, with 
    (a) $z_{med}$, (b) $z_{min}$, and (c) $\Delta z$. 
    (a--c) are for classes.
    (d--f) are for subclasses. }
    \label{fig:year_vs_hit_patent_probability}
  \end{figure}

Panels (a) and (d) of Fig.~\ref{fig:year_vs_hit_patent_probability} show that high $z_{med}$ has a small but positive influence on future impact, and vice versa for low $z_{med}$, indicating that inventions that are primarily based on established prior work do marginally better in the future.
Meanwhile, Fig.~\ref{fig:year_vs_hit_patent_probability} (b) and (e) shows that a high $z_{min}$, signifying more typical, has a noticeable negative effect on a patents future. Thus if all the pairings of an invention become conventional, it is less likely to be influential.
On the other hand, it is evident that when measuring against core conventionality \textit{and} a novel element together, the results are both more consistent and more significant, as seen in Fig.~\ref{fig:year_vs_hit_patent_probability} (c) and (f). The high $\Delta z$ has a clear positive influence, whereas the mid $\Delta z$ has no influence and the low $\Delta z$ has negative influence, capturing that is it neither of the two aspects on there own that have the most influence but the combination of the two.

We can further elaborate on this through a differing classification of the patent set: whether (i) $z_{med}$ of a patent is above or below the quartile z-score within the entire period and (ii) $z_{min}$ of a patent is above or below quartile z-score within the entire period. These classifications directly capture (i) whether the patent has a conventional core and (ii) whether it includes a novel aspect. We also redefine high impact inventions as being in the top 5th across \textit{all} the patent records (1976--2001). These criteria split patents into one of four categories: 
(high $z_{min}$, high $z_{med}$) being those patents with a conventional core but without a novel addition, namely marginal improving; (high $z_{min}$, low $z_{med}$) as neither having a conventional core nor a relatively novel addition; (low $z_{min}$, high $z_{med}$) as those patents with the success signifier of a conventional core and a novel twist; and lastly (low $z_{min}$, and low $z_{med}$) as those patents which are entirely novel, or oddball.

\begin{figure}[h!]
    \includegraphics[width=0.9\columnwidth]{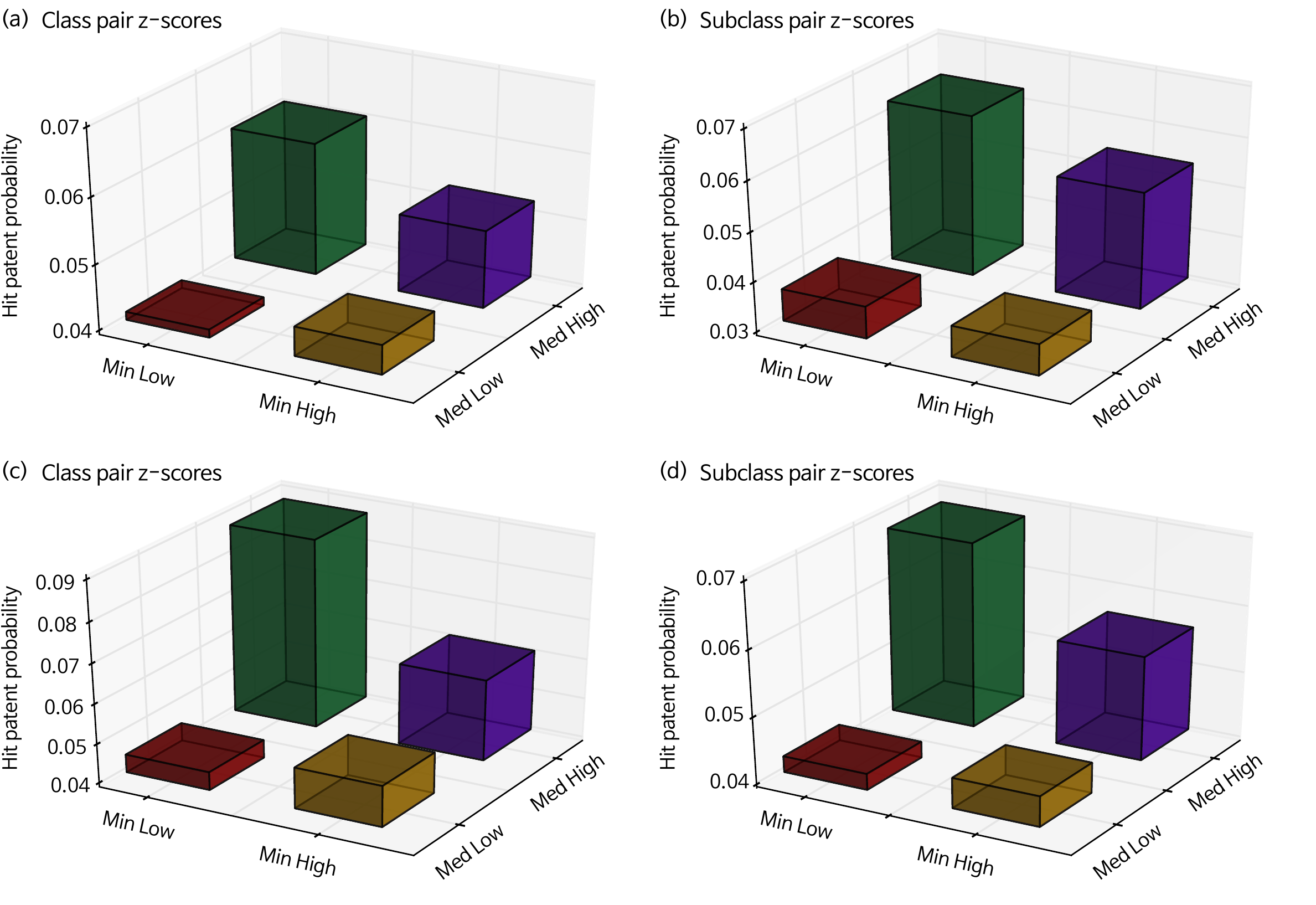}
    \caption{\textbf{The probability of a ``hit'' patent (top 5th percentile), conditional on four categories.} 
    The red bar is for low \zmed and low \zmin; the green is for high \zmed and low \zmin; the gold for low \zmed and high \zmin; 
    and the purple for high \zmed and low \zmin. 
    The thresholds of {\it high} and {\it low} are the quartile for (a) and (b); 
    the 95th percentile of \zmed and the 15th percentile of \zmin for (c); and 
    the 50th percentile of \zmed and the quartile of \zmin for (d).
    The fraction of patents in each resulted category are: 
    (a) 17.66\% (red), 7.33\% (green), 7.33\% (gold), and 67.66\% (purple);
    (b)  16.77\% (red), 8.22\% (green), 8.22\% (gold), and 66.77\% (purple); 
    (c) 14.9\% (red), 0.1\% (green), 80.1\% (gold), and 4.9\% (purple);
    and (d) 22.47\% (red), 2.5\% (green), 27.53\% (gold), and 47.47\% (purple). 
    The left (right) column are for class (subclass) pairs. }
    \label{fig:types_hit_patent_probability}
  \end{figure}

Figure~\ref{fig:types_hit_patent_probability} (a) and (b) show the ``hit'' patent probability of a patent throughout the period depending on the four categories. 
Instead of quartile that was used for (a) and (b), $z_{min}$ and $z_{med}$ are now chosen to optimise the highest hit patent probability. As a result, we can achieve the maximum probability up to almost 9\% shown in Fig.~\ref{fig:types_hit_patent_probability}, 
the full extent of the influence of the categories, with an almost doubling over the background hit patent rate for (c).
The results corroborate those in Fig.~\ref{fig:year_vs_hit_patent_probability} (c) and (f); those patents with conventional cores and a novel addition do notable better then the background rate, 
and do best of all four categories. Also, it is again shown that entirely novel inventions fair the worst.
These analyses also suggest that conventionality does not collide over novelty, but conventionality \textit{illuminated} by novelty can help an invention's influence. This finding, that is there is an optimal balance between conventionality and novelty for influential inventions, indicates that influence is associated with knowledge transfer between technological domains~\cite{Uzzi2013,Malva2014,Strumsky2015,Fleming2001}. 

\section*{5 Conclusion} 

In this paper, we quantitatively study the distribution of technology pairings' novelty and
a connection between a novelty profile in a combination and its future impact 
by using technology code pairings in the U.S. patent spanning 179 years (1836--2014)~\cite{USPTObulk}. 
We show inventions assemble technological units in a way to reinforce the already conventional pairs, 
thereby some components become increasingly entrenched within the inventive repertoire with increasing z-scores, 
such that they become a further building block for future combinations. Yet still combinations will occasionally 
bridge between these code-cliques, exhibited as increasingly negative z-scores in time. 
This result implies that the technology space forms units of tightly co-occurring codes with occasional 
inter-unit combinations to change that structure, and that inventors always require components which are 
familiar to them, or available in the industry~\cite{Arthur2011,Kauffman1996,Arthur2006,Thurner2010,Klimek2012}.

We also show how technological composition can effect the future impact of an invention, 
by associating the patents' citation count as a measure of that impact~\cite{Hall2001}.
Through analysis of citation relationships across the U.S. patents (1976--2006), 
our analysis shows the statistically significant technology pairings are correlated with future influence of an invention.
In line with the previous research, our findings demonstrate that conventional combinations, 
enlightened by proper novelty, are more likely to be influential in future, alluding to
that there is an optimal balance between conventionality and novelty for influential inventions, and
that influence is associated with knowledge transfer between technological domains~\cite{Uzzi2013,Malva2014,Strumsky2015,Fleming2001}.

Yet there still remains much research to be resolved to more rigorously quantify statistical significance of code pairings.
The proposed null model for calculating z-scores of code pairings is limited: it does not account 
for inventions having a single code and for when a code is created.
In addition, excluding citations to outside the data or academic papers may miss 
the important role of scientific research in guiding inventors to search the technological
space more efficiently, hence resulting highly novel content~\cite{Fleming2004, Koh2008, McNerney2011}. 
Future studies need to correctly complement the above limitations. 
Nonetheless, our study may provide valuable insights into how technology combinations 
give rise to boundary-spanning breakthroughs in technology as well as science and how innovative technology combinations become influential.

\section*{Competing interests}
  The authors declare that they have no competing interests.

\section*{Author's contributions}
  DK, DC and HY participated in all methodological decisions. DK collected and preprocessed the data, and simulation. DK, DC and HY analysed the result. The manuscript was written mainly by DK, DC and HY. All authors read and agreed on the final version.

\section*{Acknowledgements}
  DK, DC and HY acknowledge the support of the National Science Foundation (no. SMA-1312294).  
  This work was supported by the National Research Foundation of Korea(NRF) grant funded by the Korea government (No. 2011-0028908) (DK).  
  This work was supported by the National Research Foundation of Korea Grant funded by the Korean Government (NRF-2012-S1A3A-2033860) (HJ).
  DK, DC and HY acknowledge the support of Institute for New Economic Thinking at Oxford Martin School, Santa Fe Institute, and CABDyN. 
  The authors thank James McNerney and Fran\c{c}ois Lafond for helpful comments on an earlier version of the paper.

\bibliography{KCJY_2015}

\clearpage

\end{document}